\documentclass[useAMS,usenatbib]{mn2e}

\usepackage[dvips]{graphicx}
\usepackage{url}
\def\ii{\'{\i}}
\def\etal{{\it et~al.}}

\title[Granular physics in low-gravity environments]{Granular physics in low-gravity environments using DEM}
\author[G. Tancredi \etal]{G. Tancredi $^{1,2}$\thanks{E-mail: gonzalo@fisica.edu.uy}, A. Maciel $^{1}$, L. Heredia $^{3}$, P. Richeri $^{3}$, S. Nesmachnow $^{3}$\\
$^{1}$Departamento de Astronom\ii a, Facultad de Ciencias, Igu\'a 4225, 11400 Montevideo, URUGUAY\\
$^{2}$Observatorio Astron\'omico Los Molinos, Ministerio de Educaci\'on y Cultura, Montevideo, URUGUAY\\
$^{3}$Centro de C\'alculo, Instituto de Computaci\'on, Facultad de Ingenier\ii a, Montevideo, URUGUAY\\}
\begin{document}

\date{Accepted 2011 November 24.  Received 2011 November 21; in original form 2011 August 12}

\pagerange{\pageref{firstpage}--\pageref{lastpage}} \pubyear{2011}

\maketitle

\label{firstpage}

\begin{abstract}
Granular materials of different sizes are present on
the surface of several atmosphere-less Solar System bodies. 
The phenomena related to granular materials have been
studied in the framework of the discipline called Granular Physics; that has been studied experimentally in the laboratory and, in the last decades, by performing numerical simulations. The Discrete Element Method simulates the mechanical behavior of a media
formed by a set of particles which interact through their contact points.

The difficulty in reproducing vacuum and low-gravity
environments makes numerical simulations the most promising technique
in the study of granular media under these conditions. 

In this work, relevant processes in minor bodies of the Solar System
are studied using the Discrete Element Method. Results of simulations
of size segregation in low-gravity environments in the cases of the asteroids Eros and Itokawa are presented. The segregation of particles with different densities was
analysed, in particular, the case of comet P/Hartley 2. The surface
shaking in these different gravity environments could produce the
ejection of particles from the surface at very low relative velocities.
The shaking causing the above processes is due to: impacts, explosions
like the release of energy by the liberation of internal stresses
or the re accommodation of material. Simulations of the passage of
impact-induced seismic waves through a granular medium were also performed.

We present several applications of the Discrete Element 
Methods for the study of the physical evolution of agglomerates of
rocks under low-gravity environments. 
\end{abstract}


\section{Introduction}

Granular materials of different sizes are present on
the surface of several atmosphere-less Solar System bodies. The presence
of very fine particles on the surface of the Moon, the so-called regolith,
was confirmed by the Apollo astronauts. From
polarimetric observations and phase angle curves, it is possible to
indirectly infer the presence of fine particles on the surface of
asteroids and planetary satellites. More recently, the visit of spacecraft
to several asteroids and comets has provided us with close pictures
of the surface, where particles of a wide size range from $cm$ to hundreds
of meters have been directly observed. The presence of even finer particles
on the visited bodies can also be inferred from image analysis.

It has been proposed that several typical processes of granular materials, such as the size segregation of boulders on Itokawa, the displacement of boulders on Eros, among others (see e.g. \cite{Asphaug} and references therein), can explain some features observed on the surfaces of these bodies.
The conditions at the surface and the interior of these small Solar
System bodies are very different compared to the conditions on the
Earth's surface. Below we point out some of these differences:

\begin{itemize}
\item while on the Earth's surface the acceleration of gravity
is $9.8\ m/s^{2}$ with minor variations, on the surface of elongated
$km$-size asteroid is on the order of $10^{-2}$ to $10^{-4}\ m/s^{2}$, with
typical variations of a factor of 2
\item the presence of an atmosphere or any other fluid media
plays an important role on the evolution of grains, particularly in
the small ones (\cite{Pak}). Under vacuum conditions in space, this effect does
not occur.
\item In vacuum and low gravity conditions, other forces
might play a role comparable to that of gravity, e.g. van der
Waals forces (\cite{Scheeres}), although these forces are not considered in our present approach.
\end{itemize}

The phenomena related to granular material have been
studied in the discipline called Granular Physics. Granular media
are formed by a set of macroscopic objects (grains) which interact
through temporal or permanent contacts. The range of materials studied
by Granular Physics is very broad: rocks, sands, talc, natural and artificial powders, pills, etc.

Granular materials show a variety of behaviours under different
circumstances: when excited (fluidised), they often resemble a liquid,
as is the case of grains flowing through pipes; or they may behave
like a solid, like in a dune or a heap of sand.

These processes have been studied experimentally in
the laboratory, and, in the last decades, by numerical analysis. The numerical
simulation of the evolution of granular materials has been done recently
with the Discrete Element Method (DEM). DEM is a family of numerical
methods for computing the motion of a large number of particles such
as molecules or grains under given physical laws. DEMs simulate the mechanical behavior in a media formed by a set of particles which interact through their contact points. 

Low-gravity environments in space are difficult
to reproduce in a ground-based laboratory; especially if one
is interested in keeping a stable value of the acceleration of gravity
on the order of $10^{-2}$ to $10^{-4}\ m/s^{2}$ for several hours, since
under these low-gravity conditions the dynamical processes are much
slower than on Earth. Parabolic flights are not suitable for these
experiments, since it is not possible to attain a stable value during
the free-fall flight. For laboratory experiments, we are then left
with experiences to be carried on board space stations.

Therefore, numerical simulation is the most promising
technique to study the phenomena affecting granular material in vacuum
and low-gravity environments.

The rest of the article is organised as follows. In Section \ref{secdem} we describe the implementation of the
Discrete Element Methods used in our simulations. In Section \ref{secbne} we present the results of simulations of the process of size segregation in low-gravity environments, the so-called Brazil nut effect, in the
cases of Eros, Itokawa and P/Hartley 2. In Section \ref{secden}, the segregation
of particles with different densities is analysed, with the application
to the case of P/Hartley 2. The surface shaking in these different
gravity environments could produce the ejection of particles from
the surface at very low relative velocities; this issue is discussed in
Section \ref{secparlif}. The shaking that causes the above processes is due to impacts
or explosions like the release of energy by the liberation of internal
stresses or the re accommodation of material. Although DEM methods
are not suitable to reproduce the impact event, we are able to make
simulations of the passage of impact-induced seismic waves through
a granular media; these experiments are shown in Section \ref{secimp}. The conclusions and the applications of these results are discussed in Section \ref{secdis}.

\section{Discrete Element Methods}\label{secdem}

DEM are a set of numerical calculations
based on statistical mechanic methods. This technique is used to
describe the movements of a large amount of particles which are subjected
to certain physical interactions.

DEMs present the following basic properties that generally define this class of numerical algorithms:
\begin{itemize}
\item The quantities are calculated at points fixed to the
material. DEM is a case of a Lagrangian numerical method.
\item The particles can move independently or they can have
bounds, and they interact in the contact zones through different types
of physical laws.
\item Each particle is considered a rigid body, subject
to the laws of rigid body mechanics.
\end{itemize}
The forces acting on a particle are calculated from
the interaction of this particle with its nearest neighbors, i.e.
the particles it touches. Several types of forces are usually considered
in the literature; e.g free elastic forces, bonded elastic forces,
frictional forces, viscoelastic forces, interaction of the particles
with other objects, such as walls and mesh objects acting as boundary
conditions, global force fields (i.e. gravity), velocity dependent
damping, etc.

The main drawback of the
method is the computational cost of computing the interacting forces
for each particle at each time step. A simple all-to-all approach would require
to perform $O(N(N-1)/2)$ operations per time step, where $N$ is
the number of particles in the simulation. Several efficient methods to reduce the number
of pairs to compute have been implemented; e.g. the Verlet lists method,
the link cells algorithm, and the lattice algorithm. Another problem
for the simulation is the length of the time step, which should be much less than the
duration of the collisions, typically $1/10$ to $1/20$ of collisions duration.
Based on the Hertzian elastic contact theory, the duration of contact
($\tau$) can be expressed as: 
\begin{equation}
\tau=5.84\left(\frac{\rho(1-\nu^{2})}{E}\right)^{0.4}rv^{-0.2}
\end{equation}

(\cite{Wada}, after \cite{Timoshenko}), where $\rho$ is the grain density, $nu$ is the Poisson ratio, $E$ is the material
strength, $r$ is the radius of the particle, and $v$ is the collisional velocity.

\begin{figure}
\centering
\includegraphics[width=0.35\textwidth]{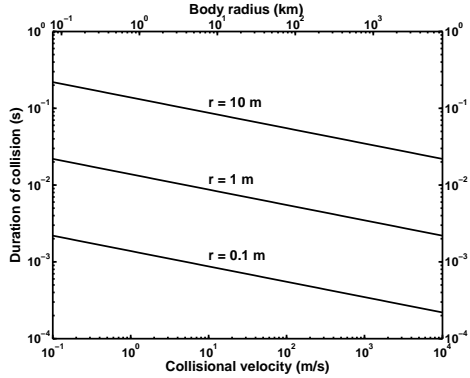}
\caption{Estimate of the duration of collision as a function of the collisional velocity for particles of $r$ = 0.1, 1 and 10 m}.
\label{fdurcol}
\end{figure}

In Figure \ref{fdurcol} we plot the previous estimate of the duration
of collision as a function of the collisional velocity for particles
with $r = 0.1$, 1 and 10 m. The other parameters are assumed as follows:
$\rho=2000\ gr/cm^{3}$, $\nu=0.17$, and $E=100\ Gpa$. We are interested
in the processes that occur on the surface of the small Solar System
bodies, where the interactions among the boulders occur at velocities
comparable to the escape velocity on their surface. The upper $x$-axis
indicates the radius of the body, while the lower one shows the corresponding
escape velocity (assuming a constant density of $\rho=3000\ gr/cm^{3}$).
For $km$-size asteroids and $m$-size boulders, the collisions typically
last for a few $10^{-3}s$, therefore, the time step required to correctly
simulate the collision would be $\sim10^{-4}s$.

\subsection{Viscolelastic spheres with friction}

The contact force between two spherical particles can
be decomposed in two vectors (Figure \ref{fcontact}): the normal force, along the
direction that joins the centres of the interacting particles; and
the tangential force, perpendicular to this line. Naming $i$ and
$j$ the two interacting particles, the total force $\overrightarrow{F_{ij}}$
can then be expressed as:

\begin{figure}
\centering
\includegraphics[width=0.3\textwidth]{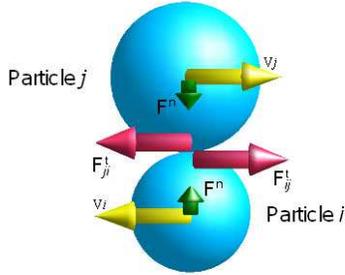}
\caption{Scheme of the contact forces between two spherical particles}
\label{fcontact}
\end{figure}

\begin{equation}
\overrightarrow{F_{ij}}=\left\{ \begin{array}{rl}
\overrightarrow{F_{ij}^{n}}+\overrightarrow{F_{ij}^{t}} & \mbox{ if \ensuremath{\psi_{ij}>0}}\\
0 & \mbox{ otherwise}
\end{array}\right.
\label{eqtotfor}
\end{equation}

where $\overrightarrow{F_{ij}^{n}}$ is the normal force and $\overrightarrow{F_{ij}^{t}}$
the tangential one. $\psi_{ij}$ is the deformation given by:

\begin{equation}
\psi_{ij}=R_{i}+R_{j}-|\overrightarrow{r_{i}}-\overrightarrow{r_{j}}|
\label{eqdefor}
\end{equation}

where $R_{i}$ and $R_{j}$ are the radius of the particle
$i$ and $j$, respectively, and $\overrightarrow{r_{i}}$ and $\overrightarrow{r_{j}}$
are the position vectors.

Several models have been used for the normal and tangential
forces in the literature. Among the most used ones is the damped dash-pot, also known as the Kelvin-Voigt model. Instead of using this model in our simulation, we use an extension of an elastic-spheres one developed by \cite{Hertz}, since it is a more realistic representation of two colliding particles.

The normal interaction force between two elastic spheres $F_{ij}^{n;el}$ was inferred by Hertz as a function of the deformation $\psi$:

\begin{equation}
F^{n;el}=\frac{2Y\sqrt{R_{eff}}}{3(1-\nu^{2})}\psi^{3/2}
\label{eqfnel}
\end{equation}

where $Y$ is the Young modulus and $\nu$ is the Poisson ratio. The effective radius $R_{eff}$ is given by the expression:

\begin{equation}
\frac{1}{R_{eff}}=\frac{1}{R_{i}}+\frac{1}{R_{j}}
\end{equation}

A viscoelastic interaction between the particles can
be modelled by including a dissipation factor in eq. \ref{eqfnel}. The viscoelastic
normal forces $F^{n;ve}$ then become: 
\begin{equation}
F^{n;ve}=\frac{2Y\sqrt{R_{eff}}}{3(1-\nu^{2})}\left(\psi^{3/2}+A\sqrt{\psi}\frac{d\psi}{dt}\right)
\label{eqfnve}
\end{equation}

where $A$ is a dissipative constant and $d\psi/dt$
is the time derivative of the deformation.

Considering the previous expression for the normal force
could lead to unrealistic results, since it does not take into account
the fact that the particles do not overlap, but they become deformed
(\cite{Poschel}). During the compression phase and most of the decompression
phase, the term $\left(\psi^{3/2}+A\sqrt{\psi}\frac{d\psi}{dt}\right)$
in eq. \ref{eqfnve} is positive, leading to a repulsive (positive) normal force.
However, at a certain stage of the decompression, the deformation
$\psi$ could still be positive, but the second term could be negative,
which would lead to a negative (attractive) force. This is an unrealistic
situation, since there are no attractive forces during the collision
of two particles. The problem arises when the centres of the particles
separate too fast from one another to allow their surfaces to keep
in touch while recovering their shape. In order to overcome this problem,
for the condition $\psi>0$ in eq. \ref{eqtotfor}, we use the following
expression for the normal force:

\begin{equation}
F^{n;ve}=\max\left\{ 0,\frac{2Y\sqrt{R_{eff}}}{3(1-\nu^{2})}\left(\psi^{3/2}+A\sqrt{\psi}\frac{d\psi}{dt}\right)\right\}
\label{eqfnvecor} 
\end{equation}

Following the model by \cite{Cundall} for the tangential
force ($F^{t}$), when two particles first touch, a shear spring is
created at the contact point. The static friction is then modelled
as a spring acting in a direction tangential to the contact plane.
The particles start sliding with the shear spring resisting the motion.
When the shear force exceeds the normal force multiplied by the friction
coefficient, dynamic sliding starts. We limit the shear
force by Coulomb's friction law; i.e. $|F^{t}\leq\mu F^{n;ve}|$.
The expression for the tangential force then becomes:

\begin{equation}
F^{t}=-sign(v_{rel}^{t})\ \min\{\|\kappa\varsigma\|\ ,\ \mu\|F^{n;ve}\|\}
\label{eqft}
\end{equation}

where the first term inside the curly brackets corresponds
to the static friction, and the second one is the dynamic friction.
$\kappa$ is a constant, $\varsigma$ is the elongation of the spring,
and $\mu$ is the dynamic friction parameter.

\subsection{ESyS-particle}

For the DEM simulations we developed a version of the ESyS-particle package (\cite{Abe2004}; https://launchpad.net/esys-particle) adapted to our needs. ESyS-particle is an Open Source software for particle-based
numerical modeling, designed for execution on parallel supercomputers, clusters or multi-core computers running a Linux-based operating system. The C++ simulation engine implements a spatial domain decomposition for parallel programming via the Message Passing Interface (MPI). A Python wrapper API provides flexibility in the design of numerical models, specification of modeling parameters and contact logic, and analysis of simulation data.

The separation of the pre-processing, simulation and post-processing tools facilitates the ESyS-particle development and
maintenance. The setup of the model geometry is given by scripts, since the whole package is script driven (no interactive GUI is provided by ESyS).

The particles can be either rotational or non-rotational
spheres. The material properties of the simulated solids can be elastic,
viscoelastic, brittle or frictional. Particles can be bonded to other
particles in order to simulate breakable material. It is possible
to implement triangular meshes for specifying boundary conditions
and walls. The package also includes a variety of particle-particle
and particle-wall interaction laws; such as linear elastic repulsion
between unbounded contacting particles, linear elastic bonds between
bonded particle pairs, non-rotational and rotational frictional interactions
between unbounded particles, rotational bonds implementing torsion
and bending stiffness and normal and shear stiffness. Boundary
conditions and walls can move according to pre-defined laws.

The DEM implementation in ESyS-particle employs the explicit integration approach, i.e. the calculation of the state of
the model at a given time only considers data from the state of the model at earlier times. Although the explicit approach requires shorter time steps, it is easier to develop a parallel version for execution in high performance computing infrastructures.

ESyS-particle has shown good scaling performance
when using additional computing elements (processor cores), if at least $\sim 5000$ particles are processed by each core. Otherwise, when a lower number of particles is handled by each core, the impact of the overhead by the communications between processes reduces the computational efficiency of the application. As long as the problem size
is scaled with the number of cores, the scalability is close to linear. Therefore, large amounts of particles, typically a few million, are possible to model.

For the analysis of the results, ESyS-particle can
format the output to be used in 3D visualisation platforms like VTK
and POV-Ray. In particular, we use the software Paraview, based on
VTK and developed by Kitware Inc. and Sandia National Labs (EEUU),
which offers good quality in 3D graphics and allows us to implement
several visualisation filters to the data.

ESyS-particle has been employed to simulate earthquake
nucleation, comminution in shear cells, silo flow, rock fragmentation,
and fault gouge evolution, to name but a few applications. Just to
give a few references, we mention examples in fracture mechanics (\cite{Schopfer}),
fault mechanics (\cite{Abe2005}, \cite{Mair}), and fault rupture
propagation (\cite{Abe2006}).

For our simulations, we have implemented the Hertzian
viscoelastic interaction model with and without friction into the ESyS-particle package,
according to the eqs. \ref{eqfnvecor} and \ref{eqft}. Several modifications were necessary to implement either in the C++ code as well as in the Python interface (\cite{Heredia}). 

\subsection{Tests}

In order to test the code and to set the values of
the relevant physical and simulation parameters, we choose a few problems
for which there exists an analytical solution or we can compare the
output with experiments.

\subsubsection{Test case 1: a direct collision of two equal spheres}

We consider the case of two equal viscoelastic spheres:
one starts at rest and the other one approaches from the negative
$x$-direction at a given speed along the line joining the particle
centres. Friction is not considered, since the collision between the
particles is normal. The aim of this test is to study the viscoelastic
collision.

The coefficient of restitution can be used to characterise
the change of relative velocity of inelastically colliding particles.
Let us note $\overrightarrow{v_{1}}$ and $\overrightarrow{v_{2}}$
the velocities before the collision of particles 1 and 2, respectively;
and $\overrightarrow{v'_{1}}$ and $\overrightarrow{v'_{2}}$ the
velocities immediately after. When the relative velocity is along
the line joining the particle centres, we note $v=|\overrightarrow{v_{2}}-\overrightarrow{v_{1}}|$
and $v'=|\overrightarrow{v'_{2}}-\overrightarrow{v'_{1}}|$. The coefficient
of restitution $\epsilon$ is then calculated as: 
\begin{equation}
\epsilon=\frac{v'}{v}
\end{equation}

In general, this coefficient depends not only on the
impact velocity, but also on material properties. 

Because of their deformation, particles lose contact
slightly before the distance of the centres between the spheres reaches
the sum of the radii. \cite{Schwager} present an analytical estimate
of the coefficient of restitution which takes into account this fact.
The computation of $\epsilon$ is then presented as a divergent series
of the dimensionless parameter $\beta v^{1/5}$, where $\beta=\gamma\kappa^{-3/5}$;
$\gamma=\frac{3}{2}\frac{\rho A}{m_{eff}}$; $\kappa=\frac{\delta}{m_{eff}}$;
$\delta=\frac{2Y}{3(1-\nu^{2})\sqrt{(}R_{eff}}$; $\frac{1}{R_{eff}}=\frac{1}{R_{i}}+\frac{1}{R_{j}}$;
and $\frac{1}{m_{eff}}=\frac{1}{m_{i}}+\frac{1}{m_{j}}$. The material
parameters $Y$, $\nu$ and $A$ are defined above. $R_{1}$, $R_{2}$,
$m_{1}$, $m_{2}$ are the radius and mass of particle 1 and 2, respectively.

We run simulations of two colliding particles with
the following combination of parameters: $Y=\{10^{9},10^{10}\}\ Pa$,
$A=\{10^{-4},10^{-3}\}\ s^{-1}$, $\nu=0.3$; for a set of initial relative
velocities $v=\{0.1,0.5,1,5,10\}\ m/s$. The particles have a radii
of 1m and a density of $3000\ kg/m^{3}$. The time step of the integration
is $10^{-5}\ s$.

\begin{figure}
\centering
\includegraphics[width=0.5\textwidth]{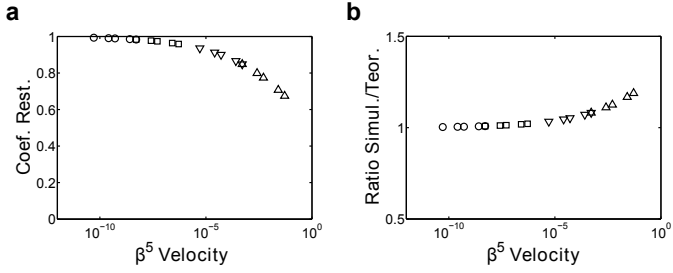}
\caption{a) The coefficient of restitution for the numerical simulations of the collision of two viscoelastic spheres as a function of $\beta v^{1/5}$. The symbols correspond to different combinations of parameters: circle \textendash \ $Y=10^{9}$, $A=10^{-4}$; down triangle \textendash \ $Y=10^{9}$, $A=10^{-3}$; square \textendash \ $Y=10^{10}$, $A=10^{-4}$; up triangle \textendash \ $Y=10^{10}$, $A=10^{-3}$ ($Y$ in $Pa$ and $A$ in $s^{-1}$). b) The ratio between the numerical and analytical estimate.}
\label{fcoefres}
\end{figure}

The coefficient of restitution for the numerical simulations
is presented in Figure \ref{fcoefres}a as a function of $\beta v^{1/5}$. The symbols correspond to different combinations of parameters: circle \textendash \ $Y=10^{9}$, $A=10^{-4}$; down triangle \textendash \ $Y=10^{9}$, $A=10^{-3}$; square \textendash \ $Y=10^{10}$, $A=10^{-4}$; up triangle \textendash \ $Y=10^{10}$, $A=10^{-3}$ ($Y$ in $Pa$ and $A$ in $s^{-1}$).
The analytical estimates are computed with Maple's codes presented
in \cite{Schwager}, where the expansions are up to 40th
order. The ratio between the numerical and analytical estimate is
presented in Figure \ref{fcoefres}b. We find a good agreement between the two estimates
up to values of $\beta v^{1/5}$ closer to 1. The discrepancy is due
to the cut-off of the higher terms.

Several laboratory experiments have been conducted
to estimate the coefficient of restitution of rock materials (see
e.g. \cite{Imre}, \cite{Durda}). For impact velocities in
the range $1-2\ \ m/s$, values of $\epsilon\sim0.8-0.9$ have been
obtained. Looking back at Figure \ref{fcoefres}a, we observe that this range of
values of $\epsilon$ are obtained for the following set of material
parameters: $Y=10^{10}\ Pa$, $A=10^{-3}\ s^{-1}$, $\nu=0.3$. Therefore,
we will choose these parameter values for our numerical simulations
of colliding rocky spheres.

\subsubsection{Test case 2: a grazing collision between two spheres}

In this case we consider two equal viscoelastic spheres:
one starts at rest and the other one approaches from the negative
$x$-direction at a given speed; but, in contrast to the previous
case, the distance between the $y$-values of the particles centres is slightly less than the sum of the radius. We then have a grazing collision. The aim of this test is to compare the results of the viscoelastic
interaction with and without friction. We run simulations of two colliding
particles with the following set of parameters: $Y=10^{10}\ Pa$,
$A=10^{-3}\ s^{-1}$, $\nu=0.3$, $R_{1}=R_{2}=1\ m$, and a density
of $\rho=3000\ kg/m^{3}$. The friction parameters of eq. \ref{eqft} are chosen
as: $\kappa=0.4$, $\mu=0.6$. The distance between the centres in
the $y$-direction is $(0.999R_{1}+R_{2})$. We run simulations where
particle 2 has initial velocities of $v=\{10^{-3},0.01,0.1,1,10\}\ \ m/s$.

\begin{figure}
\centering
\includegraphics[width=0.4\textwidth]{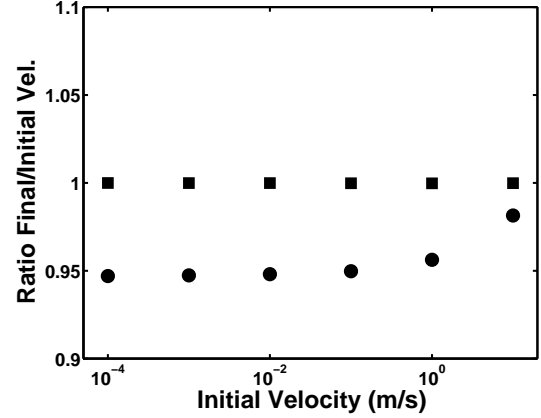}
\caption{The ratio between the modulus of the relative velocity between the particles
after exiting and before the interaction as a function of the initial velocity. The symbols correspond to different contact force models: square \textendash \ Hertzian viscoelastic spheres; circle \textendash \ Hertzian viscoelastic spheres with friction.}
\label{fratiocol}
\end{figure}

In Figure \ref{fratiocol} we plot the ratio between the modulus of
the particles relative velocity after exiting and before the interaction,
as a function of the initial velocity. The star symbols correspond
to the simulations without the friction interaction and the cross
symbols to the ones with it. Due to the fact that the collision is
almost grazing, the ratios are almost 1 for the simulations without
friction, regardless of the initial velocity. For simulations with
friction, as it is expected, the ratio decreases as the initial velocity
decreases, because the friction interaction becomes more relevant
for low velocities.

\subsubsection{Test case 3: a bouncing ball}

In ESyS-particle the interaction between a particle
and a mesh wall can be linear, elastic or a linear elastic bond. Viscoelastic
and frictional interactions of particles and walls are not yet implemented.
Therefore, in order to simulate a frictional viscoelastic collision
of a ball against a fixed wall, we have to glue balls to the wall
with a linear elastic bond. The following test case consists on a
free-falling ball impacting on an equal size ball that is bonded to
the floor. The objective of this experiment is to test different time steps
for the simulations.

We use the following set of parameters: $Y=10^{10}\ Pa$,
$A=10^{-3}\ s^{-1}$, $\nu=0.3$, $R_{1}=R_{2}=1\ m$, and a density
of $\rho=3000\ kg/m^{3}$. The bonded particle has an elastic bond
with a modulus $K=10^{9}\ Pa$. Particle 2 falls from a height of
$2.75m$. In the first set of simulations we assume the Earth's surface
gravity ($g=9.81\ \ m/s^{2}$). For this set, we use the following
time steps in the simulations: $dt=\{6\times10^{-4},5\times10^{-4},10^{-4},10^{-5},10^{-6},5\times10^{-7}\}\ s$.

The duration of the collisions is computed from the
simulations as the interval of time while the deformation parameter defined in eq. \ref{eqdefor} is greater than 0. As mentioned above this
interval is slightly larger than the time the balls are in contact,
but it is good enough for the purpose of having an order of magnitude
estimate of it. For the previous set of parameters, the duration of
the collision is $\sim0.003\ s$.

\begin{figure}
\centering
\includegraphics[width=0.5\textwidth]{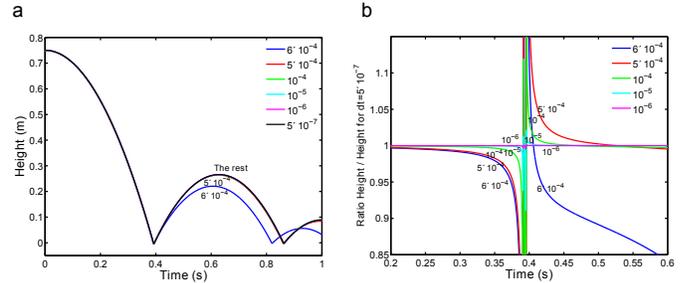}
\caption{a) The distance of the falling particle respect to the edge of the resting one (centre height minus $3R$) as a function of time. b) The ratio of the previous values to the one for the smallest time step at each time.}
\label{frelheight}
\end{figure}

In Figure \ref{frelheight}a, we plot the distance of the falling particle respect to the edge of the resting one (centre height minus $3R$) as a function of time. In Figure \ref{frelheight}b, we plot the ratio of the previous values to the one for the smallest time step at each time. We find that for time steps $dt \le 10^{-5}\ s$, there is a very good agreement between the simulations.
For longer time steps the bouncing ball presents an implausible behavior.

The coefficient of restitution can be computed as the ratio between the velocity at the iteration step just after the collision and at the step just before the collision (just after and before the deformation defined in eq. \ref{eqdefor} is $\psi < 0$). For time steps $dt \le 10^{-5}\ s$ there is a good agreement among the different estimates. We obtained a value of 0.593.

Therefore, for the previous set of parameters, we will
use a time step of $dt=10^{-5}\ s$ for the simulations with Earth's
gravity, since the collision is covered with $\sim$30 time steps and
it is a good compromise between quality of the results and a longer
time step.

In another set of simulations we use very low surface gravity, similar to the one found on the surface of asteroid Itokawa and comet P/Hartley 2; i.e. a rocky object of $\sim500\ m$ in diameter
or an icy object of $\sim1\ km$ in diameter. For this set, we use
the following time steps in the simulations: $dt=\{5\times10^{-4},10^{-4},10^{-5},10^{-6}\}\ s$.
For the previous set of parameters, the duration of the collision
is $\sim0.01\ s$. For time steps $dt\le10^{-4}\ s$ there is a good
agreement among the different runs. The coefficient of restitution
in these simulations is 0.721. For the simulations in this low-gravity
environments we will use a time step $dt=10^{-4}\ s$, which corresponds
to a collision lasting $\sim100$ time steps.

\subsubsection{Test case 4: Newton's cradle}

A Newton's cradle is a device used to demonstrate the
conservation of linear momentum and energy via a series of swinging
hard spheres. When one ball at the end is lifted and released, it
knocks a second ball and this one the next until the last ball in
the line is pushed upward. A typical Newton's cradle consists of a
series of identically sized metal balls hanging by equal length strings
from a metal frame so that they are just touching each other at rest.

We simulate the Newton's cradle with four spheres aligned
in the x-axis. We number the particles from right to left: \#1 being the particle at the
right extreme and \#4 the one at the left extreme. The x-axis increases to the right. Particle \#1 has a negative initial velocity $v_{x}=-10\ \ m/s$.
Two types of simulation are run: Hertzian elastic and viscoelastic
spheres. We use the following set of parameters: $Y=10^{10}\ Pa$,
$A=10^{-3}\ s^{-1}$, $\nu=0.3$ (for the viscoelastic simulation).
The radius of the spheres are $R=1\ m$, and a density of $\rho=3000\ kg/m^{3}$.

\begin{figure}
\centering
\includegraphics[width=0.5\textwidth]{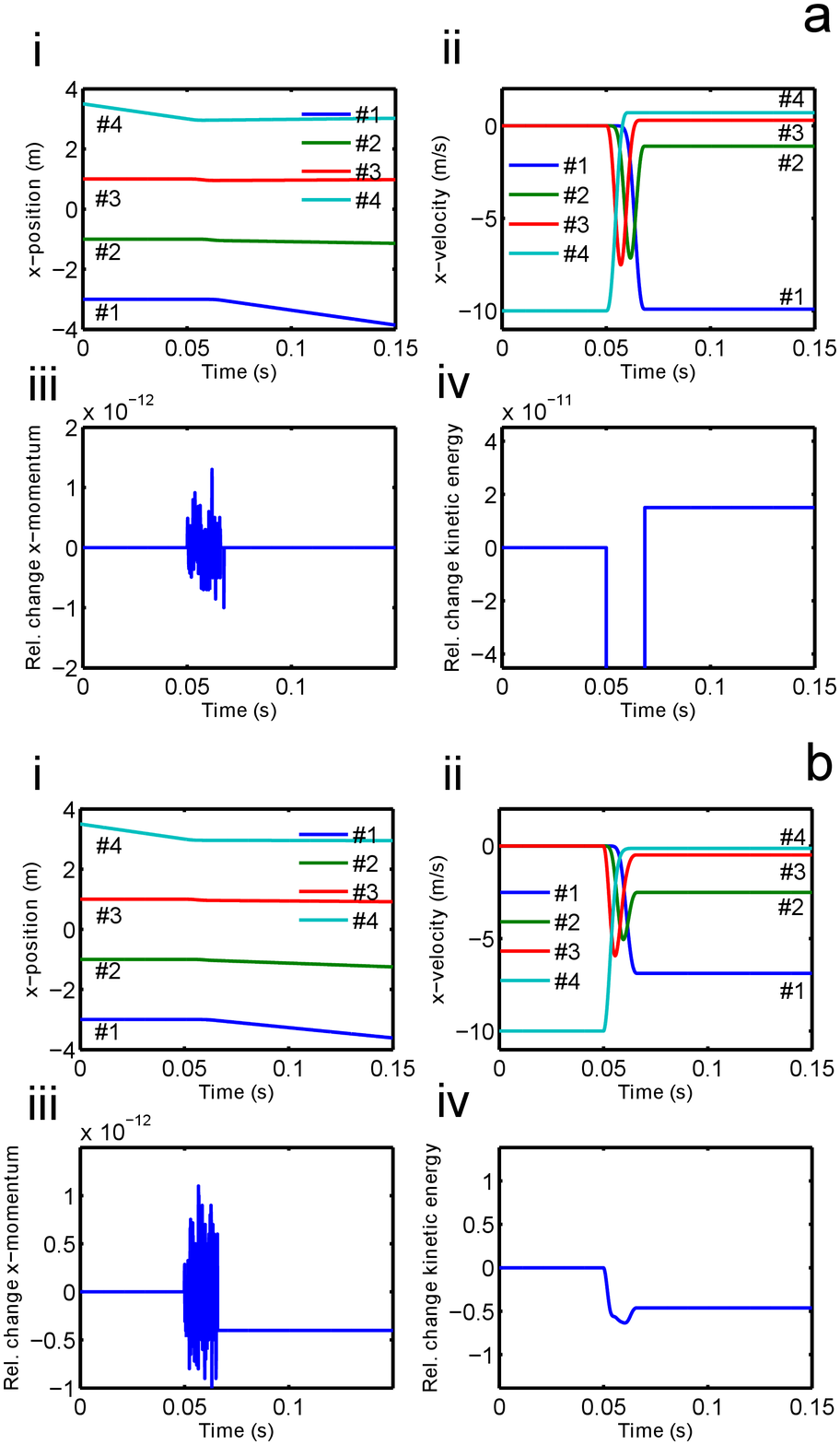}
\caption{a) Results of the Herztian elastic (HE) simulations of the Newton's cradle: i)
$x$-position of each particle (\#1 to \#4); ii) $x$-velocity for each particle; iii) relative change of $x$-total momentum; iv) relative change of the total kinetic energy. b) Similar set of plots for the Herztian viscoelastic (HVE) simulations of the Newton's cradle.}
\label{fnewton}
\end{figure}

In Figure \ref{fnewton} we present the time evolution of the following
parameters for each simulation: i) $x$-position of each particle (\#1 to \#4);
ii) $x$-velocity for each particle; iii) relative change of $x$-total
momentum: $(Momentum(t)-Momentum(t=0))/Momentum(t=0)$; iv) relative
change of total kinetic energy: $(K.E(t)-K.E.(t=0))/K.E.(t=0)$. Figure
\ref{fnewton} a) corresponds to the Herztian elastic (HE) simulation, and Figure
\ref{fnewton} b) to the Herztian viscoelastic (HVE) one.

Note that for the HE simulation particle \#4
acquires almost the velocity of the initial impacting particle and
little rebound is observed in the particles \#1 to \#3. The linear
momentum is conserved after the collision up to a relative precision
$<10^{-12}$, and the kinetic energy after the rebound is conserved
up to a relative precision of $10^{-11}$. In the HVE simulation,
the particle \#4 acquires 70\% of the velocity of the initial impacting
particle, and particle \#3 acquires 25\%. No rebound is observed and
all the particles move to the left. The final velocities increase
from right to left. The linear momentum is
also conserved after the collision up to a relative precision $<10^{-12}$
(down to the last output digit). The kinetic energy after the rebound
is not conserved $\sim$ 50\% of the initial kinetic energy is spent
on the damping of the viscosity interaction.

\section{Size segregation in low-gravity environments: The Brazil
nut effect}\label{secbne}

\subsection{The shaking or knocking procedure}

Consider a recipient with one large ball on the bottom
and a number of smaller ones on top of it. All the balls have similar
densities. After shaking the recipient for a while, the large ball
rises to the top and the small ones sink to the bottom (\cite{Rosato},
\cite{Knight}, \cite{Kudrolli}). This is the so called Brazil nut effect (BNE),
because it can be easily seen when one mixes nuts of different sizes
in a can; the large Brazil nuts rise to the top of the can. Unless
there is a large difference in the density of the balls, a mixture
of different particles will segregate by size when shaken.

The BNE has been attributed to the following processes (\cite{Hong}): i) the percolation effect, where the smaller ones pass through the holes created by the larger ones (\cite{Jullien}); ii) geometrical reorganisation, through which small particles readily fill small openings below the large particles (\cite{Rosato}); iii) global convection which brings the large particles up but does not allow for reentry in the downstream (\cite{Knight}); iv) due to its larger kinetic energy, the large particle still follows a ballistic upraise, penetrating by inertia into the bed (\cite{Nahmad-Molinari}).

While size ratio is a dominant factor, particle-specific properties such as density, inelasticity and friction can also play important roles.

\cite{Williams} performed a model experiment with a single large particle (intruder) and a set of smaller beads inside
a rectangular container. When the container was vibrated appropriately, the intruder would always rise and reach a height in the bed that depends on vibration strength.

In order to simulate this effect under different gravity conditions, we run simulations of a 3D box with many small particles and one big particle at the bottom, the so-called intruder model system (\cite{Williams}, \cite{Kudrolli}). On the floor we glue one row of small particles with a linear elastic bond. The box is subjected to a given surface gravity.

We run simulations under several gravity conditions: the surface of the Earth, Moon, Ceres, Eros and a very-low gravity
environment like the surface of asteroid Itokawa or comet P/Hartley 2. The parameters for the simulations are summarised in Table \ref{tabsur}. The physical and elastic parameters of the particles are similar to the ones used in the previous tests: $Y=10^{10}\ Pa$, $A=10^{-3}\ s^{-1}$, $\nu=0.3$, $\kappa=0.4$, $\mu=0.6$, $K=10^{9}\ Pa$, $\rho=3000\ kg/m^{3}$.

\begin{figure}
\centering
\includegraphics[width=0.35\textwidth]{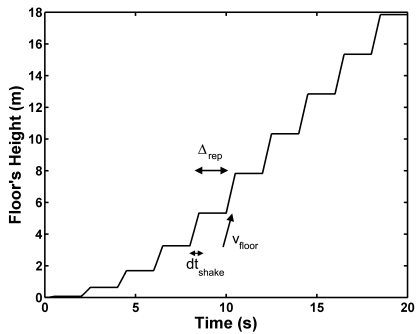}
\caption{The floor is vertically displaced at a certain speed ($v_{floor}$) for a short interval ($dt_{shake}$), according to a staircase-like function.}
\label{fstair}
\end{figure}

The floor is vertically displaced at a certain speed
($v_{floor}$) for a short interval ($dt_{shake}$), according to
a staircase-like function like the one presented in Fig \ref{fstair}. The process
is repeated every given number of seconds ($\Delta t_{rep}$), depending
on the settling time given by the surface gravity. We have chosen
this vibration scheme instead of the frequently used sinusoidal oscillation
of the floor, because we are interested in the effects of a sudden
shock coming from below. This shock could arise from the translation
of the impulse generated by an impact in a far region. We refer this vibration scheme as a shaking or knocking procedure.

In order to prepare the initial conditions for the simulations of the BNE, we run a set of simulations where the
particles start at a certain height over the surface and they free fall. The floor is slightly shaken at the beginning of these preliminary simulations in order to obtain a random settling of the particles. After finishing the shaking and letting the particles settle down, we use the positions at the end of the runs as the initial conditions
for the set of BNE simulations. We must run different preliminary simulations for each gravity environment.

In the BNE simulations, the floor's velocity is linearly increased from 0 up to the final value $v_{floor}$, which is reached after 20 jumps. We note that the shaking procedure is parameterized with the floor's velocity.

The 3D box is constructed with elastic mesh walls.
The box has a base of $6\times6\ m$ and a height of 150 $m$. A set
of $12\times12\ m$ small balls of radius $R_{1}=0.25\ m$ are glued
to the floor. The big ball has a radius $R_{2}=0.75\ m$, and on top
of it, there are 1000 small balls with a normal distribution of radii
(mean radius $R_{1}=0.25\ m$, standard deviation $\sigma=0.01\ m$).
We use the same box for all the simulations.

The size range of the balls are selected in correspondence with the boulders size observed on the surface of asteroid Itokawa and Eros.

\begin{table*}
\caption{Parameters for the simulations of the BNE under different gravity environments.}
\begin{tabular}{|l|r|r|r|r|r|}
\hline 
Parameter  & Earth  & Moon  & Ceres  & Eros  & Low-gravity\tabularnewline
 &  &  &  &  & Itokawa \& \tabularnewline
 &  &  &  &  & P/Hartley 2\tabularnewline
\hline 
\textit{\small {Surface gravity $g\ (m/s^{2})$ }} & 9.81  & 1.62  & 0.27  & $5.9\times10^{-3}$  & $10^{-4}$\tabularnewline
\textit{\small {Escape velocity $v_{esc}\ (m/s)$ }} & $11.2\times10^{3}$  & $2.38\times10^{3}$  & 510  & 10  & 0.17 \tabularnewline
\textit{\small {Floor's velocity $v_{floor}\ (m/s)$ }} & 0.3 - 10  & 0.1 - 3  & 0.03 - 1  & 0.01 - 0.3  & 0.003 - 0.1\tabularnewline
\textit{\small {Duration of displacement $dt_{shake}\ (s)$ }} & 0.1  & 0.1  & 0.1  & 0.1  & 0.1\tabularnewline
\textit{\small {Time between displacements $\Delta_{rep}\ (s)$ }} & 2  & 5  & 5  & 15  & 15\tabularnewline
\hline 
\label{tabsur}
\end{tabular}
\end{table*}

\subsection{Earth}

\begin{figure}
\centering
\includegraphics[width=0.4\textwidth]{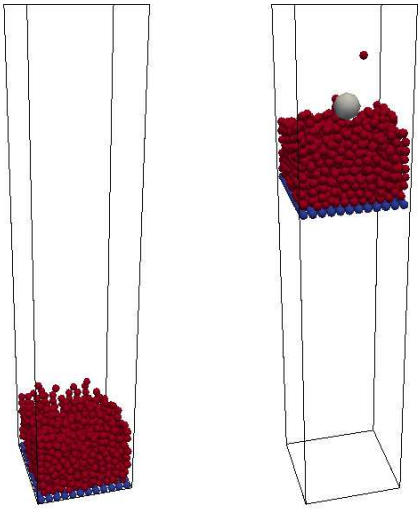}
\caption{Snapshots at start and after 50 shakes (100 sec. of simulated time) for the simulation under Earth's gravity. The snapshots correspond to the simulation with floor's velocity $v_{floor}= 5\ \ m/s$. See the movies in the supplementary material.}
\label{fsnapbneearth}
\end{figure}

We run simulations with the following set of floor
velocities: $v_{floor}=\{0.3,1,3,5,10\}\ \ m/s$. 
Snapshots at start and after 100 shakes (100 sec. of simulated time) are presented in Figure \ref{fsnapbneearth}. The snapshots correspond to the simulation with floor's velocity $v_{floor}= 5\ \ m/s$. In the supplementary material we include movies with the complete simulation (movie1 with all the spheres drawn and movie2 with the small spheres erased).

\begin{figure}
\centering
\includegraphics[width=0.4\textwidth]{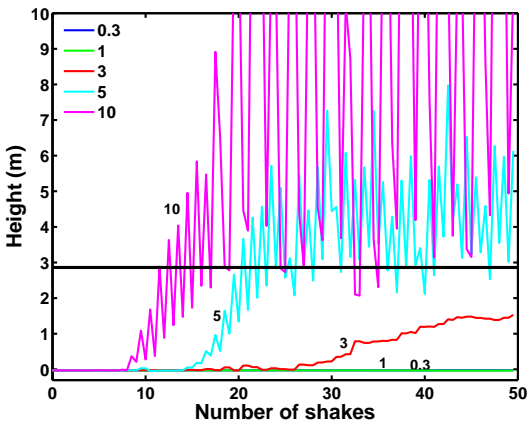}
\caption{The evolution of the big ball's height as a function of the number of shakes for different floor velocities ($v_{floor}=\{0.3,1,3,5,10\}\ \ m/s$) under Earth's gravity. Note that the floor's velocity is used as the varying parameter in the shaking process. A black line is drawn at a height of 2.84 $m$, which is the height of a compact enclosing box (see text).}
\label{fevolbigearth}
\end{figure}

In Figure \ref{fevolbigearth} we present the evolution of the big ball's height as a function of the number
of shakes for the different floor velocities. The thick black line
marks the height of a box enclosing the 1000 small particles with a random close
packing. Random close packing has a maximum porosity of $P = 0.64$ (\cite{Jaeger}).
The volume of the enclosing box is calculated as the sum of the volume
of the 1000 small particles divided by the porosity; i.e.: $V = 1000 (\frac{4/3}pi R_1^3)/P = 102\ m^3$. For a box with a $6\times6\ m$ base, we obtain a height of the enclosing box of 2.84 $m$. The thick black line is drawn at this height.

For the two lowest velocities ($v_{floor}=\{0.3,1\}\ \ m/s$)
the big ball stays at the bottom, for the two largest ones ($v_{floor}=\{5,10\}\ \ m/s$)
it rises to the top, and for the intermediate one ($v_{floor}=3\ \ m/s$)
it starts rising but does not reach the top at the end of the simulation.

When the floor's displacement velocity is below $\sim3\ \ m/s$,
the Brazil nut effect does not occur. Above this threshold, the time
required by the big ball to reach the top decreases for increasing
floor velocities. Note that there is a sharp decrease in the rising
time for small changes in the floor's velocity (from 3 to 5 $m/s$).
For large displacement velocities, the balls on the top, including
the big one that is 27 times more massive than the small ones, can
be lifted at considerable heights, as it is seen in the large excursions
made by the big ball for $v_{floor}=10\ \ m/s$.

\subsection{Comparison with other gravity environments}

Similar simulations were run for other gravity environments,
like the surface of the Moon, Ceres, Eros and a very-low gravity environment
like the surface of asteroid Itokawa or comet P/Hartley 2. The simulation
parameters are presented in Table \ref{tabsur}.

For the simulation under the very-low gravity environment, we present a movie of 4500 sec. of simulated time (300 shakes) in the supplementary material. The movie corresponds to the simulation with floor's velocity $v_{floor}= 0.05\ \ m/s$ (movie3 with all the spheres drawn and movie4 with the small spheres erased).

\begin{figure}
\centering
\includegraphics[width=0.5\textwidth]{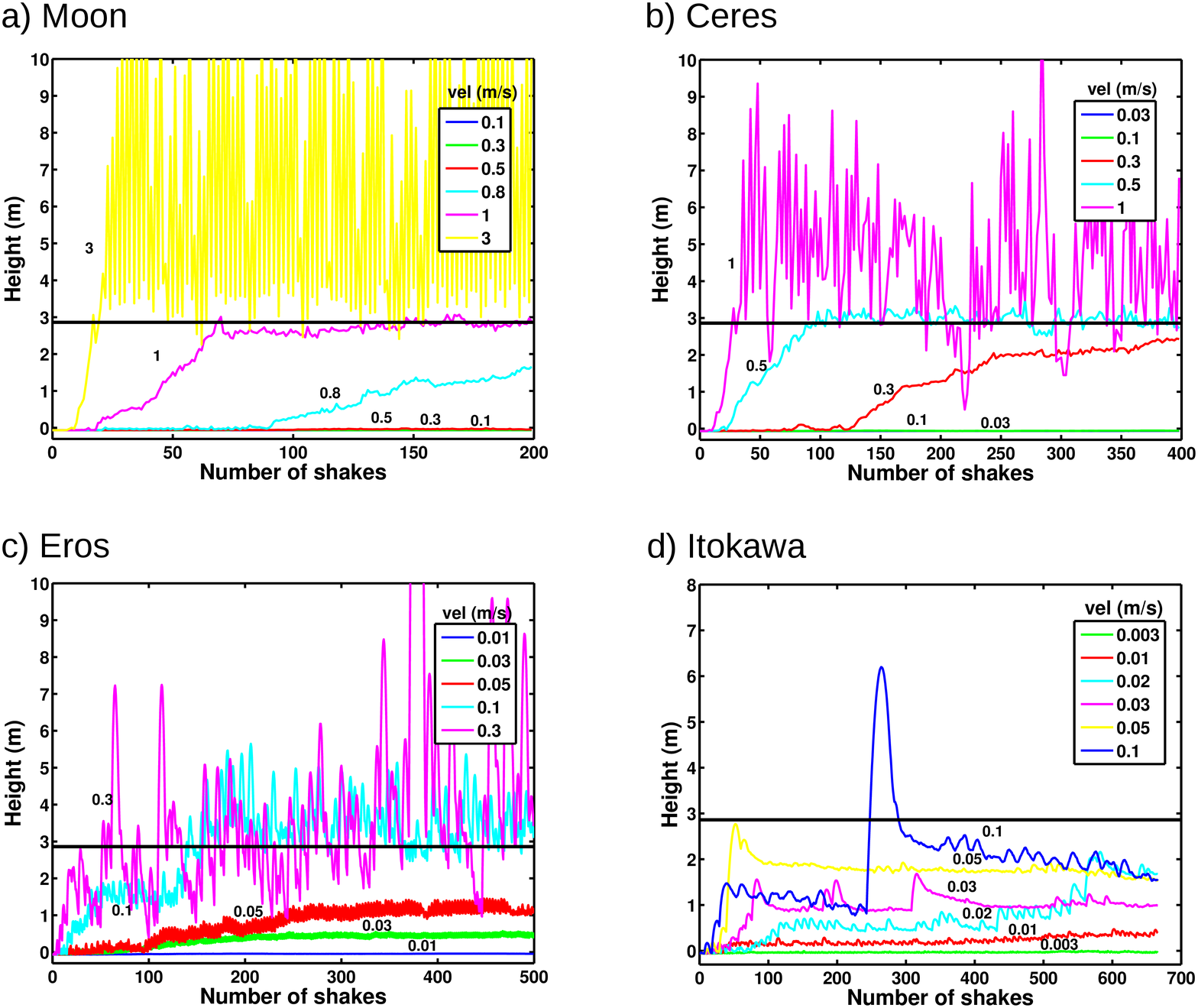}
\caption{The evolution of the big ball's height as a function of the number of shakes for different floor velocities under different gravity environments: a) Moon; b) Ceres; c) Eros; d) Itokawa. The legends correspond to the floor velocities ($v_{floor}$).}
\label{fevolbigother}
\end{figure}

Figure \ref{fevolbigother} presents the evolution of the big ball's
height as a function of the number of shakes for the different floor
velocities and the different gravity environments: a) Moon, b) Ceres,
c) Eros, d) Itokawa.

\begin{figure}
\centering
\includegraphics[width=0.4\textwidth]{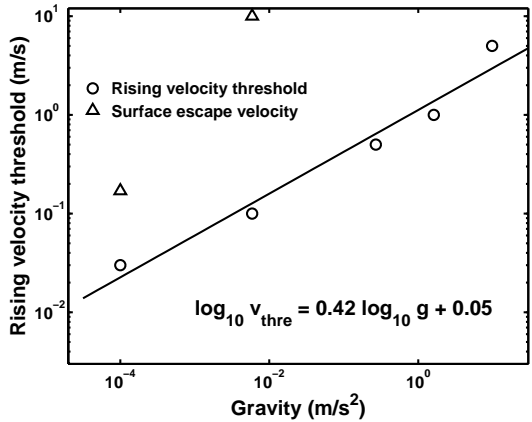}
\caption{Comparison of the floor's velocity threshold for the different gravity environments. The Brazil nut effect does not occur if the floor's velocity is below the threshold. The floor's velocity thresholds are presented as small circles. Note that the thresholds are not precisely estimate, because they are computed from the plots in Figure \ref{fevolbigearth} and \ref{fevolbigother} a-d. A straight line in the log-log space is fitted to the data points. The up triangles represent the escape velocity for the given surface gravity. The escape velocity for the largest objects are out of the plots.
}
\label{fthreshold}
\end{figure}

As in the cases of the simulations in Earth's gravity, in all the different gravity environments we can find a threshold for the floor's velocity, below which the Brazil nut effect does not occur. From the
previous plots, we get a rough estimate of these thresholds. In Figure
\ref{fthreshold} we plot the velocity thresholds as a function of the surface gravity
in a log-log scale. A straight line in the log-log space is a good
fit to the data points:

\begin{equation}
\log_{10}v_{thre}\left[\ m/s\right]=0.42\log_{10}g\left[\ m/s^{2}\right]+0.05
\end{equation}

We conclude that the Brazil nut effect is effective
in a wide range of gravity environments, expanding 5 orders of magnitude
on surface gravity.

In Figure \ref{fthreshold}, we plot the escape velocity for the given
surface gravity. Note that the floor's velocity thresholds approach
the escape velocity for the low gravity environments. For example,
in the case of Itokawa, the escape velocity is $v_{esc}=0.17\ \ m/s$,
while the estimated floor's velocity threshold is $v_{floor}=0.015\ \ m/s$.
This point is revisited in Section \ref{secparlif}.

\section{Density segregation in low-gravity environments}\label{secden}

As mentioned above, other particle-specific properties
can affect the segregation process. In particular, the effects of
density have been studied the most. For ratios of the density of the
large to the small particles much larger than 1 (denser large particles),
the segregation effect could be reversed, and the large particles
would sinks to the bottom, producing the so-called Reverse Brazil
Nut Effect (RBNE) (\cite{Shinbrot}, \cite{Hong}).

However, for particles of similar sizes but different
densities, both laboratory (\cite{Mobius}, \cite{Shi}) or numerical (\cite{Lim})
experiments have shown that the lighter particles tend to rise and
form a pure layer on the top of the system, while the heavier particles
and some of the lighter ones stay at the bottom and form a mixed layer.
In the Solar System, we might encounter bodies with such a mixture
of heavy and light particles. Cometary nuclei are believed to be formed
of a mix of icy and rocky material. However, the intimacy of this
mixture is still unknown, with two possible scenarios: 1) every particle
is made of a mixture of ice and dust, and 2) there exist some particles
mainly formed by icy material and some others mainly formed by rocky
constituents that are mixed together. 

We shall investigate the behavior of a mixture of light
and heavy particles under different gravity environments.

For the simulations we create a 3D box similar to the
previous one, with a $6\times6\ m$ base and a height of 150 $m$.
The box is constructed with elastic mesh walls. On the floor we glue
a set of $12\times12$ small balls of radius $R_{1}=0.25\ m$ and
density $\rho=2000\ kg/m^{3}$. There are 500 light balls with a normal
distribution of radii (mean radius $R_{1}=0.25\ m$, standard deviation
$\sigma=0.01\ m$) and density $\rho=500\ kg/m^{3}$. On top of them,
there are 500 heavy balls with the same distribution of radii and
density $\rho=2000\ kg/m^{3}$. At the beginning of the simulations
the balls are placed sparsely, the light balls at the bottom and the
heavy ones on top. They free fall and settle down before starting
the floor shaking.

Elastic parameters of the particles are the same for
both types of particles and similar to the ones used in the previous
tests for all the particles: $Y=10^{10}\ Pa$, $A=10^{-3}\ s^{-1}$,
$\nu=0.3$, $\kappa=0.4$, $\mu=0.6$, $K=10^{9}\ Pa$.

The floor is displaced with a staircase function in
a similar way as in the previous set of simulations.

Two gravity environments were tested: the Earth's surface
gravity and a very-low gravity environment like the surface of comet
P/Hartley 2.

\begin{figure}
\centering
\includegraphics[width=0.35\textwidth]{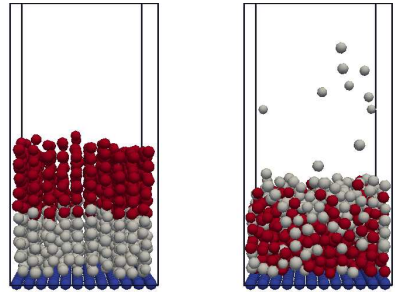}
\caption{Snapshots of the initial and final state (after
1300 shakes) for a simulation under the low-gravity environment and
a floor velocity of $v_{floor}=0.05\ \ m/s$. (see movie5 and movie6 in the supplementary material.}
\label{fsnapdensitylow}
\end{figure}

Figure \ref{fsnapdensitylow} presents snapshots of the initial and final state (after
1300 shakes) for a simulation under the low-gravity environment and
a floor velocity of $v_{floor}=0.05\ \ m/s$. In the supplementary material we include  
movies with the complete simulations (movie5 corresponds to the simulation under Earth's gravity and $v_{floor}=3\ \ m/s$; movie6 corresponds to the simulation under low gravity and $v_{floor}=0.05\ \ m/s$. Note that in these movies the camera moves with the floor, therefore it seems that the floor is always located in the same position, but it really is moving with the staircase function described above).

\begin{figure}
\centering
\includegraphics[width=0.5\textwidth]{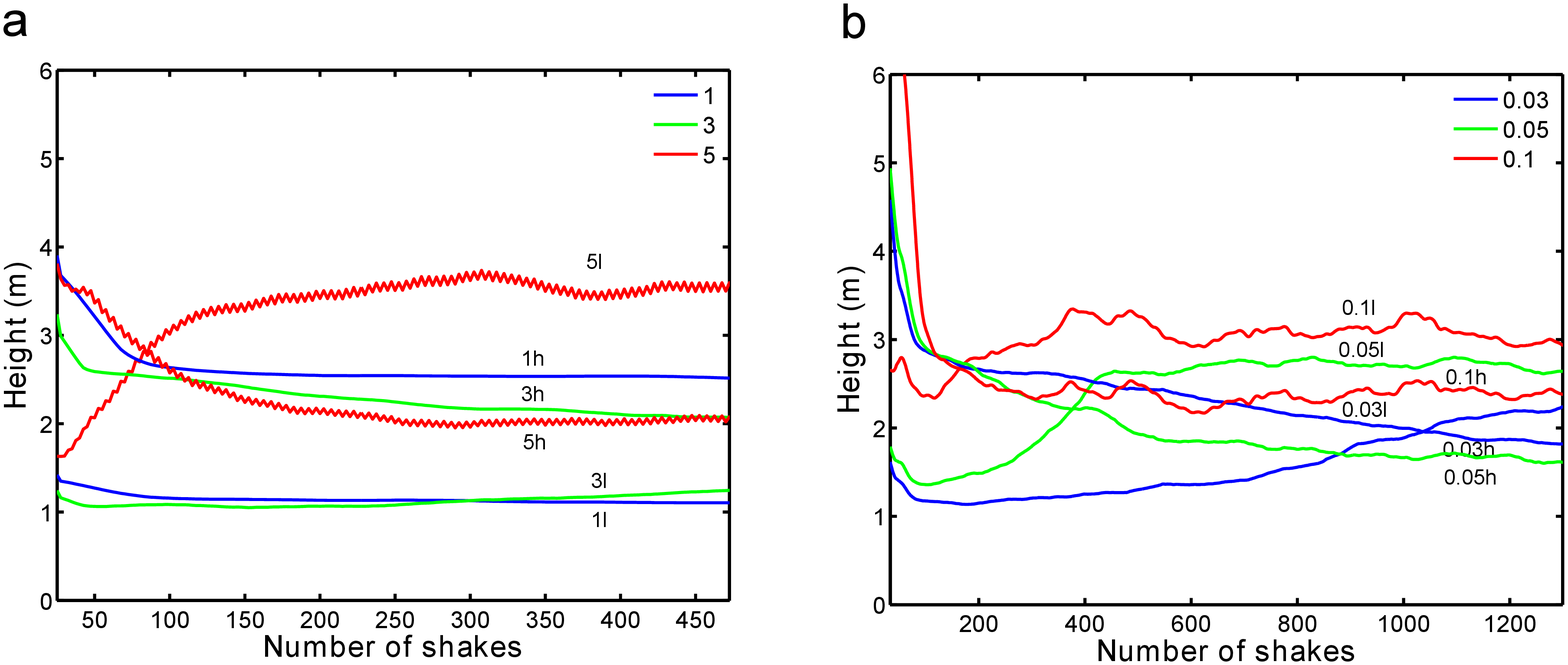}
\caption{The median height of the light and heavy particles are plotted as a function of the number of shakes for different floor velocities and under different gravity environments: a) Earth, b) low-gravity like P/Hartley 2. The legends correspond to the floor velocities ($v_{floor}$).
For each simulation there are two lines: the one that starts on top
corresponds to the medium height of the heavy particles (labeled with \textsf{h}) and the one that starts at the bottom to the medium height of the light ones (labeled with \textsf{l})}
\label{fevoldensityearth}
\end{figure}

At every snapshot, we compute the median height of
the light and heavy particles, respectively. These median heights
are plotted as a function of the number of shakes in Figure \ref{fevoldensityearth} a) for
the Earth's gravity simulations, and b) for the low-gravity ones.
For each simulation there are two lines: the one that starts on top
corresponds to the heavy particles and the one that starts at the
bottom to the light ones. In the Earth environment simulations, the
lines do not cross for the two lowest floor velocities: $v_{floor}=\{1,3\}\ \ m/s$;
therefore, the particles do not overturn the initial segregation.
Though, for $v_{floor}=3\ \ m/s$, the lines start to approach. However,
for the highest floor velocities, i.e. $v_{floor}=5\ \ m/s$, the
lines cross at an early stage of the simulation after which they remain
almost parallel. Most of the light particles move to the top and most
of the heavy ones sink to the bottom; the end state is similar to
the one seen in Figure \ref{fsnapdensitylow} for the low-gravity simulations. Due to the
strong shakes, the particles suffer large displacements, but, in a
statistical sense, the two set of particles are segregated. A density
segregation is then observed, although it is not complete.

The results of the simulations under a low-gravity
environment are presented in Figure \ref{fevoldensityearth}b. The lines for the light and
heavy particles median height do cross for the three studied floor
velocities ($v\_{floor}=\{0.03,0.05,0.1\}\ \ m/s$), although for
the lowest velocity the simulations do not last long enough to reach
the stable stage where the median heights reach almost a stable value.

Note that in both gravity environments, the density
segregation is effective for floor's velocity over a threshold similar
to the ones of the size segregation effect of Section \ref{secbne}.

\section{Particle lifting and ejection}\label{secparlif}

Let us consider the following simple experiment: we
have a layer of material that is uniformly shocked from the bottom.
The motivation of this experiment is to consider what would happen
if a seismic wave, generated somewhere in a body and propagating through
it, reaches another region of the body from below. What would happen
with material deposit on the surface? Let us take into account three different
materials: a solid block, a compressible fluid and a set of grains.
The outcome of the experiment will be different depending on the material.
When the seismic wave knocks the solid block, the block is pushed
upward. It starts to move upward, forming a gap between the
layer's bottom and the floor. In the case of a layer of compressible
fluid, an elastic p-wave is transmitted through it, producing compression
and rarefaction of the material.

But, what happens in the case of a layer of grains?
Before presenting the results of some simulations, let us reconsider
the simulations of Newton's cradle with Hertzian viscoelastic spheres.
We have seen that after the first particle knocks the second one from
the right, all the particles move to the left. Particle \#4, the last one 
on the row, moves faster, the next one to the right moves slower and
so forth. Therefore, the whole set of particles move in the same direction,
but they do not do it as a compact set, the particles separate from
each other.

We perform a first set of simulations with a homogeneous
set of particles. A 3D box with a base of $7.5\times7.5\ m$ is filled
with $15\times15=225$ particles glued to the bottom, with a radius
$R=0.25\ m$. We create 2744 particles with a mean radius $R_{1}=0.25\ m$,
standard deviation $\sigma=0.01\ m$ and density $\rho=3000\ kg/m^{3}$.
To generate the initial conditions for the simulations, these particles
are located a few cm from the bottom and they free fall under the
different gravity environments until they settle down.

Elastic parameters of all the particles are similar
to the ones used in the previous tests for all the particles: $Y=10^{10}\ Pa$,
$A=10^{-3}\ s^{-1}$, $\nu=0.3$, $\kappa=0.4$, $\mu=0.6$, $K=10^{9}\ Pa$.

\begin{figure}
\centering
\includegraphics[width=0.35\textwidth]{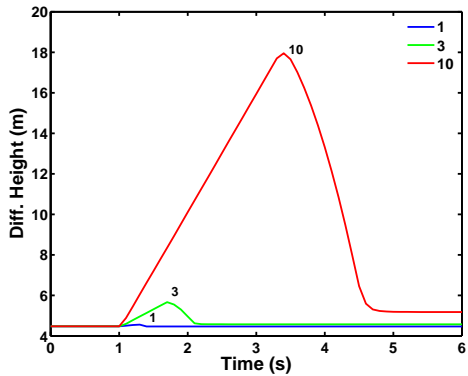}
\caption{Lifting of particles under Earth gravity. At every snapshot, we sort the particles by their height respect to the floor, and we compute the height of the 10\% ($h_{10}$) and 90\% ($h_{90}$) percentile. 
We plot the difference of these two quantities ($h_{90}-h_{10}$) for the different
floor velocities.}
\label{fdiffdensityearth}
\end{figure}

With the initial conditions generated above, we run the following
experiment: after a given time ($t_{sep}$), the floor is vertically
displaced at a certain speed ($v_{floor}$) for a short interval ($dt_{shake}$),
only one time. Two gravity environments are used for the simulations:
Earth's surface and the low-gravity environment of Itokawa. For the
Earth's simulations we use the following set of parameters: $t_{sep}=1\ s$,
$v_{floor}=\{1,3,10\}\ \ m/s$, $dt_{shake}=0.1\ s$. At every snapshot,
we sort the particles by their height respect to the floor, and we
compute the height of the particles at the 10\% ($h_{10}$) and 90\% ($h_{90}$) percentile. 
In Figure \ref{fdiffdensityearth} we plot the
difference of these two quantities ($h_{90}-h_{10}$) for the different
floor velocities. We observe that these differences increase with
time up to a certain instant when the particles fall back. Therefore,
the particles are not moving as a compact set, rather, the upper particles
are moving faster and the particles separate from each other. The
upper particles can reach velocities larger than the floor's velocity;
e.g. in the case of $v_{floor}=10\ m/s$, the 10\% fastest particles
reach velocities of $\sim17\ m/s$ just after the end of the floor's
displacement. We observe that the upper particles are lifted at considerable
heights before they fall back.

Similar results are obtained in low-gravity simulations, using the following set of parameters: $t_{sep}=10\ s$, $v_{floor}=\{0.01,0.03,0.1\}\ m/s$, $dt_{shake}=0.1\ s$. The upper particles move faster and they can reach velocities up to $\sim0.02,0.05,0.2\ m/s$ with respect to the floor velocities. Note that the escape velocity in this environment is $v_{esc}=0.17\ m/s$, therefore the fastest ejection velocities of the lifted particles are higher than $v_{esc}$. We run another experiment: on top of the layer of particles with mean radius $R_{1}\sim0.25\ m$, we deposit a layer of 2700 smaller particles, with mean radius $R_{2}=0.1\ m$ and standard deviation $\sigma=0.01\ m$. The rest of the physical
parameters are the same as for the bigger particles. The aim of this experiment is to check whether the small particles are ejected with higher velocities than the big ones. As in the previous simulations, we order the particles in increasing height. We compute the height of the 90\% percentile of the big ($h_{b,90}$) and small particles ($h_{s,90}$). Although the small particles on top of the big ones tend to separate, the differences in the velocities are relatively
small. There is no significant ejection of the small particles.

\begin{figure}
\centering
\includegraphics[width=0.4\textwidth]{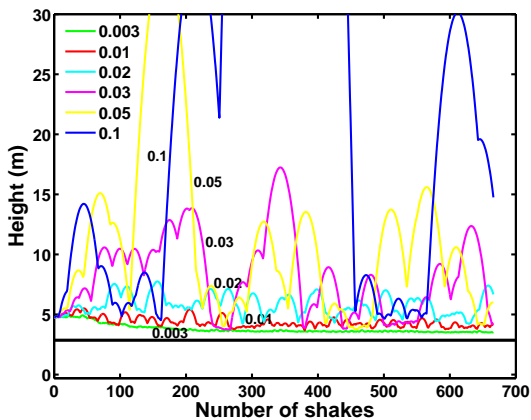}
\caption{The maximum height of the particles as a function of the simulated
time for the case of the low-gravity environment and different floor
velocities.}
\label{fheightlow}
\end{figure}

Another relevant result regarding the lifting and ejection
of particles from the surface due to an incoming shock from below,
can be obtained from the Brazil nut effect simulations presented in
Section \ref{secbne}. In the animations produced with a sequence of snapshots
for the simulations where the segregation process was effective, we
observe many particles lifted at considerable heights. In Figure \ref{fheightlow}
we plot the maximum height of the particles as a function of the simulated
time for the case of the low-gravity environment and different floor
velocities. Note that the ejection velocities the fastest particles
can acquire are comparable to the floor's displacement velocities,
and even, a little bit higher. For a floor velocity of $0.1\ m/s$,
the particles can reach an ejection velocity higher than the escape
velocity at the surface.

Taking into consideration the previous results, we
conclude that a layer shocked from below would produce the lifting
of particles at the surface if the displacement of the bottom exceeds
a certain velocity threshold. Particles can acquire vertical velocities
comparable to the displacement velocity of the bottom. For very low-gravity
environments, this velocity could be comparable to the escape velocity
at the surface. The particles could enter in sub-orbital or orbital
flights, creating a cloud of gravitational weakly bounded particles
around the object.

\section{Global shaking due to impacts and explosions}\label{secimp}

In the previous sections we have shown that several
physical processes can occur in a layer of granular media when it
is shocked from below: size and density segregation, lifting and ejection
of particles. A big quake in a distant point could produce such a
shock. The quake could be produced by another small object impacting
the body or by the release of some internal stress. Interplanetary
impacts typically occur at velocities of several $km/s$. These are
hypervelocity impacts, i.e. impacts with velocities that are above
the sound speed in the target material, which give rise to physical
deformation of the target, heating and shock waves spreading out from
the impact point. The DEM algorithms described above can not successfully
reproduce these set of phenomena. Therefore, we have to implement a
different approach if we are interested in understanding the effect
of an impact induced shock wave passing through a granular media.

Let's consider a $km$-size agglomerated body, formed
by many $m$-size boulders. We raise the following question: what happens
if a small projectile impacts in such an object at distances far from
the impact point? Or alternatively, what happens if a large amount
of kinetic energy is released in a small volume close to the surface
of such an object? In order to answer these questions we run the following
set of simulations. We fill a sphere of radius 250 and 1000 $m$ with
small spheres of a given size range, using the configurations, number of moving particles, and total
mass listed in Table \ref{tabcase}. For each sphere, we fill the volume with
two different distributions of small spheres: one with $\sim 90,000$
particles and another one with a larger number of particles $\sim 700,000$.
We try to make the total mass of the moving particles similar for
each of the studied radii.

A time step of $dt = 10^{-4} \ s$ is used in all the simulations. The simulations are run in a cluster with Intel Xeon multi-core processors (Model E5410, at 2.33 GHz, with 12MB Cache). For cases B and D we use up to 8 cores. In these cases, a simulation of 10 $s$ takes $\sim 20 \ hr$ of CPU-time in each core.

\begin{table*}
\caption{Parameters for the simulations of underground explosions}
\begin{tabular}{|l|r|r|r|r|}
\hline 
Case  & A  & B  & C  & D\tabularnewline
\hline 
Parameter  & Radius  & Radius  & Radius  & Radius\tabularnewline
 & 250 m  & 250 m  & 1000 m  & 1000 m \tabularnewline
\hline 
Size range of spheres ($m$)  & 2.5 - 12.5  & 1 - 10  & 10 - 50  & 5 - 25 \tabularnewline
Number of particles  & 88570  & 783552  & 89144  & 688443 \tabularnewline
Porosity  & 0.31  & 0.22  & 0.31  & 0.31\tabularnewline
Total Mass ($10^{12}kg$)  & 0.135  & 0.152  & 8.66  & 8.63\tabularnewline
Escape velocity at surface ($m/s$)  & 0.269  & 0.285  & 1.075  & 1.073\tabularnewline
Number of initially moving particles  & 10  & 140  & 10  & 200 \tabularnewline
Mass of moving particles ($10^{6}kg$)  & 21  & 21  & 599  & 608 \tabularnewline
\hline 
\textit{\small {Energy-equivalent projectile radius ($m$) for $v=100m/s$ }} & 9.88  & 0.88  & 2.67  & 2.68\tabularnewline
\textit{\small {Energy-equivalent projectile radius ($m$) for $v=500m/s$ }} & 2.57  & 2.56  & 7.82  & 7.85\tabularnewline
\textit{\small {Momentum-equivalent projectile radius ($m$) for
$v=100m/s$ }} & 3.26  & 3.23  & 9.84  & 9.89\tabularnewline
\textit{\small {Momentum-equivalent projectile radius ($m$) for
$v=500m/s$ }} & 5.53  & 5.52  & 16.83  & 16.92\tabularnewline
\textit{\small {Ratio Kinetic Energy / Potential Energy for $v=100\ m/s$ }} & 36  & 29  & 1  & 1 \tabularnewline
\textit{\small {Ratio Kinetic Energy / Potential Energy for $v=500\ m/s$ }} & 907  & 710  & 25  & 26 \tabularnewline
\textit{\small {Specific energy $Q^{*}$ ($J/kg$) for $v=100\ m/s$ }} & 0.79  & 0.69  & 0.35  & 0.35\tabularnewline
\textit{\small {Specific energy $Q^{*}$ ($J/kg$) for $v=500\ m/s$ }} & 20  & 17  & 8.6  & 8.8 \tabularnewline
\hline
\label{tabcase} 
\end{tabular}
\end{table*}

Since we can not successfully simulate the physics
of a hypervelocity impact during the very short initial stages, we
implemented another approach. At a given point on the surface we select
a certain number of particles of the body that are close to this place.
Each particle has at the beginning of the simulation a velocity along
the radial vector toward the centre. We substitute the impact by a
near-surface underground explosion, where several particles are released at a
given speed. For each set of configurations listed in Table \ref{tabcase}, we
run simulations with initial particle velocities of 100 $m/s$ and
500 $m/s$. These velocities are well below the sound speed in the
target material.

Since these initial conditions would correspond to
a stage after the impact where some energy has already been spent
in the compression, fracturing and heating of the target material,
we cannot equal the sum of the kinetic energy of the moving particles
with the kinetic energy of the impactor. However, we can provide a lower limit to the kinetic energy of the impactor by assuming efficiency factor $\epsilon_{KE}=1$, or a corresponding lower limit of the impactor size for a given impact velocity. In Table \ref{tabcase}, we also present the radius of the equivalent
projectile for the two set of initial particle velocities, assuming
an energy efficiency factor $\epsilon_{KE}=1$ and an impact velocity
of 5 $km/s$. For lower values of the efficiency factor, the projectile
radius would scale with $\epsilon_{KE}^{-1/3}$. As we have seen in
the simulations of the Newton's cradle with viscoelastic interactions,
there is a considerable loss of kinetic energy after a series of collisions,
although the total linear momentum is conserved. As far as we know,
there is very limited data on the transfer of momentum in hypervelocity
impacts, and we do not know the efficiency factor of this transfer
($\epsilon_{LM}$). A similar estimate of the lower limit for the impactor size can be done by assuming a momentum efficiency factor $\epsilon_{LM}=1$
and an impact velocity of 5 $km/s$. In Table \ref{tabcase}, we present the radius
of the equivalent projectile for the two set of initial particle velocities.
The projectile radius would scale with $\epsilon_{LM}^{-1/3}$.

\begin{figure}
\centering
\includegraphics[width=0.5\textwidth]{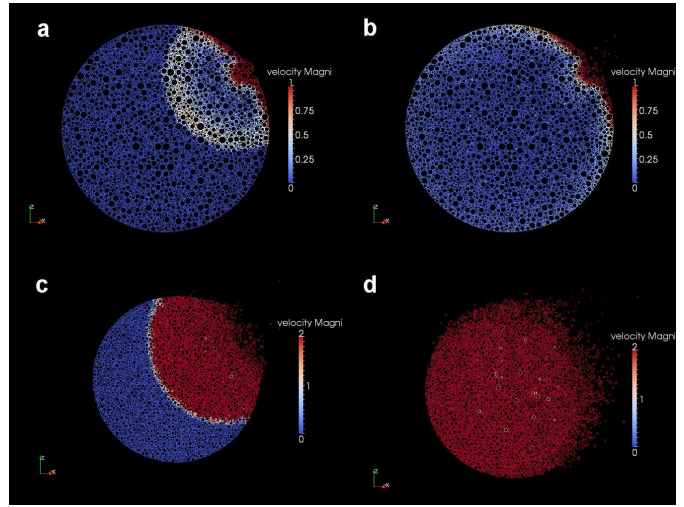}
\caption{Snapshots of the sphere explosions simulations. These are slices passing through the centre of the sphere, the explosion point and the poles. Snapshots a) and b) correspond
to the simulations with body radius of 250 $m$, the largest number
of particles ($N=783552$) and particles velocities of 100 $m/s$
(case B-100). Snapshot a is at 0.4 $s$ after the explosion and b
is at 2 $s$. The particles are coloured using a colour bar that scales
with the modulus of the velocity. Snapshots c) and d) correspond to the simulations with body radius of 1000 $m$, the largest number of particles ($N=688443$) and particles velocities
of 500 $m/s$ (case D-500). Snapshot c is at 3 $s$ after the explosion
and d is at 6 $s$. (see movies in the supplementary material)}
\label{fsnapexp}
\end{figure}

The location of the explosion is always at the surface
and with angular coordinates ($latitude=45\deg$ , $longitude=45\deg$).
In Figure \ref{fsnapexp} we present snapshots showing the propagation of the wave
into the interior, by using slices passing through the centre of
the sphere, the explosion point and the poles. Figure \ref{fsnapexp} a and b correspond
to the simulations with body radius of 250 $m$, the largest number
of particles ($N=783552$) and particles velocities of 100 $m/s$
(case B-100). Snapshot a is at 0.4 $s$ after the explosion and b
is at 2 $s$. The particles are coloured using a colour bar that scales
with the modulus of the velocity. On the other hand, Figure \ref{fsnapexp}c and
d correspond to the simulations with body radius of 1000 $m$, the
largest number of particles ($N=688443$) and particles velocities
of 500 $m/s$ (case D-500). Snapshot c is at 3 $s$ after the explosion
and d is at 6 $s$.  In the supplementary material we present movies of these simulations. (movie7 corresponds to the case B-100 $m/s$, movie8 to case B-500 $m/s$, movie9 to case D-100 $m/s$, and movie10 to case D-500 $m/s$. In the movies we observed the variation of the velocity of the particles in a slice passing through the centre of the sphere, the explosion point and the poles. The particles are coloured using a colour bar that scales
with the modulus of the velocity.

We note that a shock front with a spherical shape propagates
to the interior from the explosion point. On the surface, there appears
a layer of fast moving particles that extends until it intersects
with the spherical front, creating inside the volume limited by the
surface layer and the spherical front, a cavity of slow moving particles.
The velocity of the propagation front has a weak dependence on the
velocity of the initial particles. For example, in the simulations
of the smaller body (case B-100), the propagation shock requires 1.8
$s$ to reach the antipodes of the explosion point, implying a velocity
of 278 $m/s$. In the case B-500, the required time is 1.2 $s$, and
the velocity 416 $m/s$. For the largest body, the figures are: case
D-100: time 9.6 $s$, velocity 208 $m/s$; case D-500: time 5.8
$s$, velocity 435 $m/s$. Although there is an increase in the initial
velocity of the moving particles of a factor of 5 among the cases,
the velocity of the propagation shock has an in increase of $\sim$
2. The velocity of the propagation shock is quite constant while the
shock travels through the interior.

\begin{figure}
\centering
\includegraphics[width=0.5\textwidth]{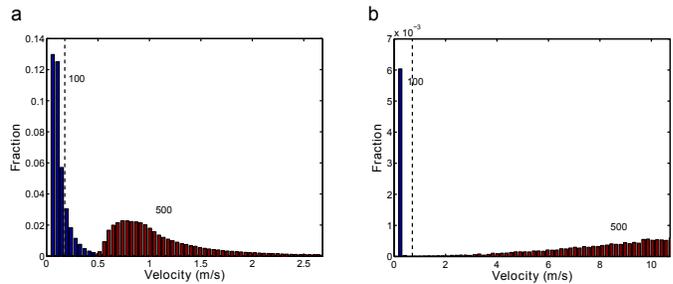}
\caption{The distribution of the ejection velocity of the particles for the simulated explosion. a) Simulations with body radius of 250 $m$ and the largest number
of particles ($N=783552$) (case B). Two histograms are presented for initial velocities of 100 and 500 $m/s$. b) Simulations with body radius of 1000 $m$ and the largest number
of particles ($N=688443$) (case D). Two histograms for initial velocities of 100 and 500 $m/s$.
In each plot, a vertical dashed line is drawn at the value of the escape velocity at the surface.}
\label{fejecvel}
\end{figure}

We are interested in the effects of the explosion at
large distances from the explosion point. The body is divided in 8
quadrants. The explosion occurs on the surface at the centre of the
first quadrant (in Cartesian coordinates the first quadrant is: $x>0\ \&\ y>0\ \&\ z>0$;
and the explosion point is at: $x=y=z=R/\sqrt{3}$, $R$ - radius).
We analyse the distribution of ejection velocities of the particles
close to the surface ($r>0.8R$) on the other 7 quadrants. Histograms
of these distributions are presented in Figure \ref{fejecvel} a and b. In Figure \ref{fejecvel}a there are two overlapping histograms which correspond to the cases
B-100 and B-500, while in Figure \ref{fejecvel}b, they correspond to the cases
D-100 and D-500. A vertical line marking the escape velocity for each
body is included in the plots. Note that for the smallest object and
for both initial velocities, there is a significant fraction of particles
that acquire ejection velocities over the escape limit. Considering
the total fraction of particles with velocities over this threshold
(not only the ones near the surface), we obtain values of 18\% in
the case B-100, and 81\% in the case B-500. In the case of the largest
body, there is a significant fraction of escaping particles only for
the largest initial velocity. The total fraction of escaping particles
are 0.6\% in the case D-100, and 100\% in the case D-500. For the
simulations with initial velocities of 500 $m/s$, there is a total
disruption of both bodies ($>50\%$ of the mass is ejected at velocities
over the escape one). It is out of the scope of this paper to derive
the disruption laws for this type of experiments; we just mention
that with a set of experiments like the previous ones, we could obtain
the kinetic energy threshold over which the explosions lead to a total
disruption of the body as a function of size. In Table \ref{tabcase} we also
include the ratio between the kinetic energy of the moving particles
over the potential energy of the body and the specific energy (defined
as the deposited energy per unit mass). \cite{Housen}
have defined the critical specific energy ($Q^{*}$) as the energy
per unit mass necessary to catastrophically disrupt a body. \cite{Ryan}
presents a plot comparing different estimates of $Q^{*}$ by several
authors as a function of the target radius. Let us note the fact that
the largest body ($R=1000\ m$) is more disrupted than the smallest
body ($R=250\ m$), although the specific energy is lower, it is in
agreement with the dip in the $Q^{*}\ vs\ R$ plot (\cite{Ryan}) in
this radius range.

Except for the case of low velocity explosions for
the large body, in all the other simulations, a fraction of the near
surface particles far from the explosion point acquire velocities
over the escape one (see Figure \ref{fejecvel}). Therefore, an explosion would induce
the ejection of particles from the surface at low velocities. These
particles could either enter into orbit around the body or slowly
escape from it, producing a cloud of fine particles that may take
many days before disappearing. This result is complementary to the
one obtained in Section \ref{secparlif} regarding the lifting and ejection of particles
produced by a shake coming from below the surface.

In the case of the smallest body, even the low velocity
explosions would induce displacement velocities over several tenths
of $m/s$ on many near surface particles far from the explosion point
(see Figure \ref{fejecvel}). This displacement would produce a shake coming from
below, similar to the shakes simulated in Section \ref{secbne}. The surface
gravity of the smallest body is similar to the surface gravity used
in the low-gravity simulations of Section \ref{secbne}, and for the largest
body the conditions are similar to the simulation of Eros. Looking
back to Figure \ref{fsnapexp}, we conclude that explosion events like the one produced
in our simulations would be enough to induce the shaking required
to produce size and density segregation on the surface of these bodies. 

This process of shaking the entire object after an impact is suitable for small bodies where the escape velocity is comparable to the impact induced displacement velocity at large distance from the impact point. Further work should study up to which body sizes the shaking process is expected to occur.

\section{Conclusions and applications of the results}\label{secdis}

The main objective of this paper is to present the applications of
Discrete Element Methods for the study of the physical evolution of
agglomerates of rocks under low-gravity environments. We have presented
some initial results regarding process like size and density segregation
due to repeated shakings or knocks, the lifting and ejections of particles from
the surface due to an incoming shock and the effect of a surface explosion
on a spherical agglomerated body. We recall that our shaking process is due to repeated set of knocks.

The main conclusions of these preliminary results are:
\begin{itemize}
\item A shaking induced size segregation \textendash the so-called
Brazil nut effect\textendash  does occur even in the low-gravity environments of the surface of small Solar Systems bodies, like $km$-size asteroids and comets.
\item A shaking induced density segregation is also observed in these environments, although it is not complete.
\item A particle layer shocked from below would produce the lifting
of particles at the surface, which can acquire vertical velocities
comparable to the surface escape velocity in very low-gravity
environments.
\item A surface explosion, like the one produced by an impact or the release of energy by the liberation of internal stresses or by the re accommodation of material, would induce a shock transmitted through the entire body, and the
ejection of surface particles at low velocities at distances far from the explosion point. This process is only suitable for small bodies.
\end{itemize}

The application of these results
to real cases will be the subject of further papers, but we foresee
some situations where the results presented here will be relevant:
\begin{itemize}
\item The internal structure of asteroid Itokawa and similar
small asteroids formed as an agglomerate of $m$-size particles, and
the relevance of the Brazil nut effect produced by repeated impacts. 
\item The non-uniform distribution of active zones in comets,
like P/Hartley 2, and the internal density segregation of icy and
rocky boulders produced by shakes caused by explosions and impacts. 
\item The formation of dust clouds at low escaping velocities
after an impact onto a $km$-size asteroid.
\end{itemize}

\bigskip

\noindent
The supplement online material can be accesed at: \\
\url{http://www.astronomia.edu.uy/Publications/}
\url{Tancredi/Granular_Physics/}

\section*{Acknowledgements}
We would like to thank Dion Weatherley, Steffen Abe and the ESyS-particle users community for helpful suggestions about the package. We thank Mariana Mart\ii nez Carlevaro for a careful reading of the text and many linguistic suggestions.

\newpage

\section*{Supplementary online material for \textit{"Granular physics in low-gravity environments using DEM"}}

Hereby you will find a set of movies included in the article "Granular physics in low-gravity environments using DEM" by Tancredi et al. (MNRAS, 2011).

The supplement online material can be accesed at: \\
\url{http://www.astronomia.edu.uy/Publications/}
\url{Tancredi/Granular_Physics/}

\section*{Size segregation (the Brazil nut effect) simulations}

A 3D box is constructed with elastic mesh walls.
The box has a base of $6\times6\ m$ and a height of 150 $m$. A set
of $12\times12\ m$ small balls of radius $R_{1}=0.25\ m$ are glued
to the floor. The big ball has a radius $R_{2}=0.75\ m$, and on top
of it, there are 1000 small balls with radii $R_{1} \sim 0.25\ m$.

The floor is displaced with a staircase function as described in the paper with different velocities. 

We present movies for two set of simulations: a) under Earth's gravity (surface gravity $g\ = 9.81 \ m/s^{2}$) and a floor's velocity ($v_{floor}= 5 \ m/s$), b) in a low-gravity environment ($g\ = 10^{-4} \ m/s^{2}$) and ($v_{floor}= 0.05 \ m/s$). 

movie1.avi is a movie with all the spheres drawn and movie2.avi with the small spheres erased for the first simulation. The movies correspond to 100 seconds of simulated time and 50 shakes.

While movie3.avi and movie4.avi correspond to the second one. The movies correspond to 10000 seconds of simulated time and 
667 shakes.

\section*{Density segregation simulations}

A 3D box similar to the previous one is created, with a $6\times6\ m$ base and a height of 150 $m$.
The box is constructed with elastic mesh walls. On the floor we glue
a set of $12\times12$ small balls of radius $R_{1}=0.25\ m$ and
density $\rho=2000\ kg/m^{3}$. There are 500 light balls with radii $R_{1} \sim 0.25\ m$ and density $\rho=500\ kg/m^{3}$. On top of them,
there are 500 heavy balls with similar radii and
density $\rho=2000\ kg/m^{3}$. At the beginning of the simulations
the balls are placed sparsely, the light balls at the bottom and the
heavy ones on top. They free fall and settle down before starting
the floor shaking.

The floor is displaced with a staircase function in
a similar way as in the previous set of simulations.

We present movies for two set of simulations: a) under Earth's gravity (surface gravity $g\ = 9.81 \ m/s^{2}$) and a floor's velocity ($v_{floor}= 3 \ m/s$), b) in a low-gravity environment ($g\ = 10^{-4} \ m/s^{2}$) and ($v_{floor}= 0.05 \ m/s$). 

movie5.avi is a movie of the first simulation, while movie6.avi corresponds to the second one. The movie5 corresponds to 1000 seconds of simulated time and 500 shakes, while the movie6 corresponds to 20000 seconds of simulated time and 1333 shakes.

Note that in these movies the camera moves with the floor, therefore it seems that the floor is always located in the same position, but it really is moving with the staircase function described above.

In movie5 the density segregation is not reached; while in movie6, most of the light particles move to the top and most of the heavy ones sink to the bottom.

\section*{Global shaking due to impacts and explosions}

We consider a km-size agglomerated body, formed
by many small size boulders. We fill a sphere of radius 250 and 1000 $m$ with
$\sim 700,000$ small spheres of a given size range (1-10 $m$-size boulders in the case of the small body, and 5-25 $m$-size boulders for the big body). 

At a given point on the surface we select a certain number of particles of the body that are close to this place.
Each particle has at the beginning of the simulation a velocity along
the radial vector toward the centre.
The location of the explosion is always at the surface
and with angular coordinates ($latitude=45\deg$ , $longitude=45\deg$).
We run simulations with initial particle velocities of 100 $m/s$ and
500 $m/s$. 

In the movies we present snapshots showing the propagation of the wave
into the interior.
These are slices passing through the centre of the sphere, the explosion point and the poles. 
The particles are coloured using a colour bar that scales with the modulus of the velocity.

movie7.avi and movie8.avi correspond to the simulation with a body of radius 250 $m$, $N=783552$ small particles and 140 particles with initial velocities of 100 $m/s$ (case B-100) and 500 $m/s$ (case B-500), respectively. 

movie9.avi and movie10.avi correspond to the simulation with a body of radius 1000 $m$, $N=688443$ small particles and 200 particles with velocities of 100 $m/s$ (case D-100) and 500 $m/s$ (case D-500). All the movies correspond to 10 seconds of simulated time.

\label{lastpage}


\begin{thebibliography}{}


\bibitem[\protect\citeauthoryear{Abe and Mair}{2005}]{Abe2005}
Abe, S., Mair, K. 2005, Geophysical Research Letters, 32, L05305, doi:10.1029/2004GL022123 

\bibitem[\protect\citeauthoryear{Abe \etal}{2004}]{Abe2004}
Abe, S., Place, D., Mora, P. 2004, Pure Appl. Geophys., 161, 2265-2277

\bibitem[\protect\citeauthoryear{Abe \etal}{2006}]{Abe2006}
Abe, S., Latham, S., Mora, P. 2006, Pure Appl. Geophys. 163, 1881–1892

\bibitem[\protect\citeauthoryear{Asphaug}{2007}]{Asphaug}
Asphaug, E. 2007, Science 316, 993-994

\bibitem[\protect\citeauthoryear{Cundall and Strak}{1979}]{Cundall}
Cundall, P., Stark, P. 1979, Geotechnique, 29, 47-65

\bibitem[\protect\citeauthoryear{Durda \etal}{2011}]{Durda}
Durda, D., Movshovitz, N., Richardson, D., Asphaug, E., Morgan, A. Rawlings, A., Vest, C. 2011, Icarus, 211, 849–855

\bibitem[\protect\citeauthoryear{Heredia and Richeri}{2009}]{Heredia}
Heredia, L., Richeri, P. 2009, Paralelismo aplicado al estudio de medios granulares, Proyecto de Grado, Inst. Computaci\'on, Fac. Ingenier\ii a, UdelaR, Uruguay

\bibitem[\protect\citeauthoryear{Hertz}{1882}]{Hertz}
Hertz, H. 1882, J. f. reine u. angewandte Math., 92, 156-171

\bibitem[\protect\citeauthoryear{Hong \etal}{2001}]{Hong}
Hong, D., Quinn, P., Luding, S. 2001, Phys. Rev. Let., 86, 3423-3426

\bibitem[\protect\citeauthoryear{Housen and Holsapple}{1990}]{Housen}
Housen, K., Holsapple, K. 1990, Icarus, 84, 226-253

\bibitem[\protect\citeauthoryear{Imre \etal}{2008}]{Imre}
Imre, B., R\"absamen, S., Springman, S. 2008, Computers \& Geosciences, 34, 339–350

\bibitem[\protect\citeauthoryear{Jaeger and Nagel}{1992}]{Jaeger}
Jaeger, H., Nagel, S. 1992, Science 255, 1523

\bibitem[\protect\citeauthoryear{Jullien and Meakin}{1992}]{Jullien}
Jullien, R., Meakin, P. 1992, Phys. Rev. Lett., 69, 640-643

\bibitem[\protect\citeauthoryear{Knight et al.}{1993}]{Knight}
Knight, J., Jaeger, H., Nagel, S. 1993, Phys. Rev. Lett., 70, 3728–31

\bibitem[\protect\citeauthoryear{Kudrolli}{2004}]{Kudrolli}
Kudrolli, A. 2004, Rep. Prog. Phys., 67, 209-247

\bibitem[\protect\citeauthoryear{Lim}{2010}]{Lim}
Lim, E. 2010, American Institute of Chemical Engineers Journal, 56, 2588-2597

\bibitem[\protect\citeauthoryear{Mair and Abe}{2008}]{Mair}
Mair, K., Abe, S. 2008, Earth and Planetary Science Letters, 274, 72–81

\bibitem[\protect\citeauthoryear{M\"obius et al.}{2001}]{Mobius}
M\"obius, M., Lauderdale, B., Nagel, S., Jaeger, H. 2001, Nature, 414, 270

\bibitem[\protect\citeauthoryear{Nahmad-Molinari et al.}{2003}]{Nahmad-Molinari}
Nahmad-Molinari, Y., Canul-Chay, G., Ruiz-Suárez, J.C. 2003, Phys. Rev. E, 68, 041301

\bibitem[\protect\citeauthoryear{Pak et al.}{1995}]{Pak}
Pak, H., Van Doom,  E., Behringer  R. 1995, Phys. Rev. Let., 74, 4643-4646

\bibitem[\protect\citeauthoryear{P\"oschel and Schwager}{2005}]{Poschel}
P\"oschel, T., Schwager, T. 2005, Computational Granular Dynamics (Springer-Verlag, Berlin Heidelberg)

\bibitem[\protect\citeauthoryear{Richardson \etal}{2005}]{Richardson}
Richardson, Jr. J., Melosh, H., Greenberg, R., O'Brien, D. 2005, Icarus, 179, 325-349.

\bibitem[\protect\citeauthoryear{Rosato et al.}{1987}]{Rosato}
Rosato, A., Strandburg, K., Prinz, F., Swendsen, R. 1987, Phys. Rev. Let., 58, 1038-1040

\bibitem[\protect\citeauthoryear{Ryan}{2000}]{Ryan}
Ryan, E. 2000, Annu. Rev. Earth Planet. Sci., 28, 367–389

\bibitem[\protect\citeauthoryear{Scheeres}{2010}]{Scheeres}
Scheeres, D. 2010, Icarus, 210, 968–984

\bibitem[\protect\citeauthoryear{Schopfer et al.}{2009}]{Schopfer}
Schopfer, M., Abe, S., Childs, C., Walsh, J. 2009, International Journal of Rock Mechanics and Mining Sciences, 46, 250--261

\bibitem[\protect\citeauthoryear{Schwager and P\"oschel}{2008}]{Schwager}
Schwager, T., P\"oschel, T. 2008, Phys. Rev. E, 78, 51304, 1-12

\bibitem[\protect\citeauthoryear{Shinbrot and Muzzio}{1998}]{Shinbrot}
Shinbrot, T., Muzzio, F. 1998, Phys. Rev. Let., 81, 4365-4368

\bibitem[\protect\citeauthoryear{Shi \etal}{2007}]{Shi}
Shi, Q., Sun, G., Hou, M., Lu, K. 2007, Phys. Rev. E, 75, 61302, 1-4

\bibitem[\protect\citeauthoryear{Timoshenko and Goodier}{1970}]{Timoshenko}
Timoshenko, S., Goodier, J. 1970, Theory of Elasticity, third ed. (McGraw–
Hill, New York)

\bibitem[\protect\citeauthoryear{Wada et al.}{2006}]{Wada}
Wada, K., Senshu, H., Matsui, T. 2006, Icarus, 180, 528–545

\bibitem[\protect\citeauthoryear{Williams}{1963}]{Williams}
Williams, J. 1963, Fuel Soc. J., 14, 29–34


\end{thebibliography}
\end{document}